\documentclass[journal]{IEEEtran}
\usepackage{cite}
\usepackage[cmex10]{amsmath}
\usepackage{graphicx}
\usepackage{color}
\usepackage{color}

\begin{document}
\bstctlcite{IEEEexample:BSTcontrol}

\title{High Resolution Flicker-Noise-Free Frequency Measurements of Weak Microwave Signals}
%
%
%

\author{Daniel~L.~Creedon,~\IEEEmembership{Student Member,~IEEE,}
        Michael~E.~Tobar,~\IEEEmembership{Fellow,~IEEE,}
        Eugene~N.~Ivanov,
        and~John~G.~Hartnett
\thanks{D. L. Creedon, M. E. Tobar, E.N. Ivanov and J. G. Hartnett are with the School of Physics, The University of Western Australia, Crawley WA 6009, Australia (e-mail: creedon@physics.uwa.edu.au). The first three authors are also with the Australian Research Council Centre of Excellence for Engineered Quantum Systems.}
}

\maketitle

\begin{abstract}
Amplification is usually necessary when measuring the frequency instability of microwave signals. In this work, we develop a flicker noise free frequency measurement system based on a common or shared amplifier. First, we show that correlated flicker phase noise can be cancelled in such a system. Then we compare the new system with the conventional by simultaneously measuring the beat frequency from two cryogenic sapphire oscillators with parts in 10$^{15}$ fractional frequency instability. We determine for low power, below $-$80 dBm, the measurements were not limited by correlated noise processes but by thermal noise of the readout amplifier. In this regime, we show that the new readout system performs as expected and at the same level as the standard system but with only half the number of amplifiers. We also show that, using a standard readout system, the next generation of cryogenic sapphire oscillators could be flicker phase noise limited when instability reaches parts in 10$^{16}$ or better.
\end{abstract}

\IEEEpeerreviewmaketitle

\section{Introduction}
%
%
%
%
\IEEEPARstart{I}{n} some applications it is necessary to amplify microwave signals to practically useful levels of power. However, it is possible that the electronic noise introduced by the amplifiers could become a concern for signals where high spectral purity is important. For example, the recent development of a cryogenic Fe$^{3+}$ sapphire maser \cite{Pyb2005apl,Benmessai2007el,Benmessai2008prl, BenmessaiAmpProc,BenmessaiGyrotropic,Creedon2010} with an output power ranging from $-$95 to $-$55 dBm requires substantial signal amplification with minimal perturbation of frequency stability. It has been documented that flicker phase noise will be induced in some amplifiers with input powers below -80 dBm \cite{IvanovAmplifiers2000}.\\

The standard method of determining the fractional frequency instability of an oscillator is to build a second nominally identical oscillator and measure the beat frequency with a high precision frequency counter to compute the frequency deviation (i.e. using the square root Allan or Triangle variance \cite{DawkinsTriangle2007}). The beat frequency is obtained with a microwave double-balanced mixer, with most mixers requiring input power on the order of 0 to +15 dBm on the RF and LO ports to operate in the correct regime.\\

It is possible that as the stability of the Fe$^{3+}$ sapphire maser is improved in future generations, nonlinearities within the amplification chain and undesirable effects such as intermodulation distortion will manifest as an artificial noise floor in the measurement. Of primary concern is amplifier-induced flicker noise, which is caused by direct, intrinsic phase modulation generated within the microwave amplifier. This is typically a low-frequency phenomenon but is of concern because it is upconverted close to the microwave carrier frequency. A readout system capable of cancelling correlated amplifier-induced flicker noise and other undesirable effects has applications for the cryogenic sapphire maser and any other system requiring amplification of very low-power signals without degrading the fractional frequency stability.\\

In this paper, we present a comparison of three microwave frequency readout systems. In our flicker-noise-free shared amplifier approach, we show that flicker phase noise is indeed correlated and largely independent of the frequency separation between the two carriers. Our readout system is important for applications that measure a difference frequency and require the signals to be amplified without the addition of extra noise. Such applications include tests of Lorentz invariance \cite{StanwixLorentz2005,StanwixLorentz2006,HohenseePRD2007}, dual mode oscillator circuits \cite{TobarMTTS2001,TobarMST2002, TobarUFFC2002,TobarMST2004,AnstieUFFC2006,TorrealbaEL2006}, and two oscillator beat frequency measurement systems, such as those presented in this paper.

\section{Amplifier Phase Noise Measurements in Dual Excitation}
\label{section2}
\begin{figure*}[!t]
\centering
\includegraphics[width=7in]{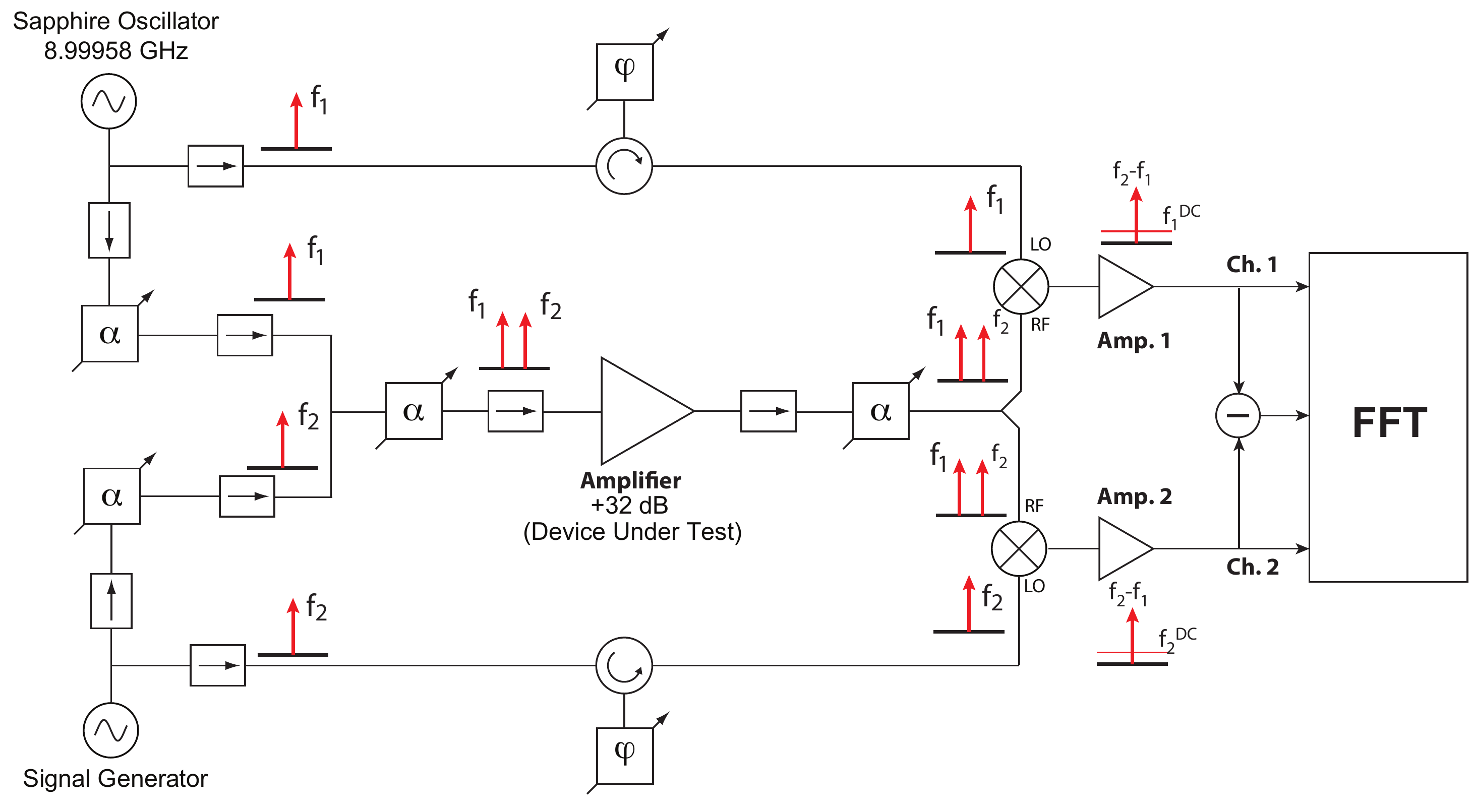}
\caption{\label{fig1}Two-channel cross correlation noise measurement system, with input frequencies $f_1$ and $f_2$. The signals $f_1^{DC}$ and $f_2^{DC}$ are the DC signals corresponding to self-mixing of $f_1$ and $f_2$ respectively. Measurements are performed in the frequency domain.}
\end{figure*}

\begin{figure}[!b]
\includegraphics[width=3.3in]{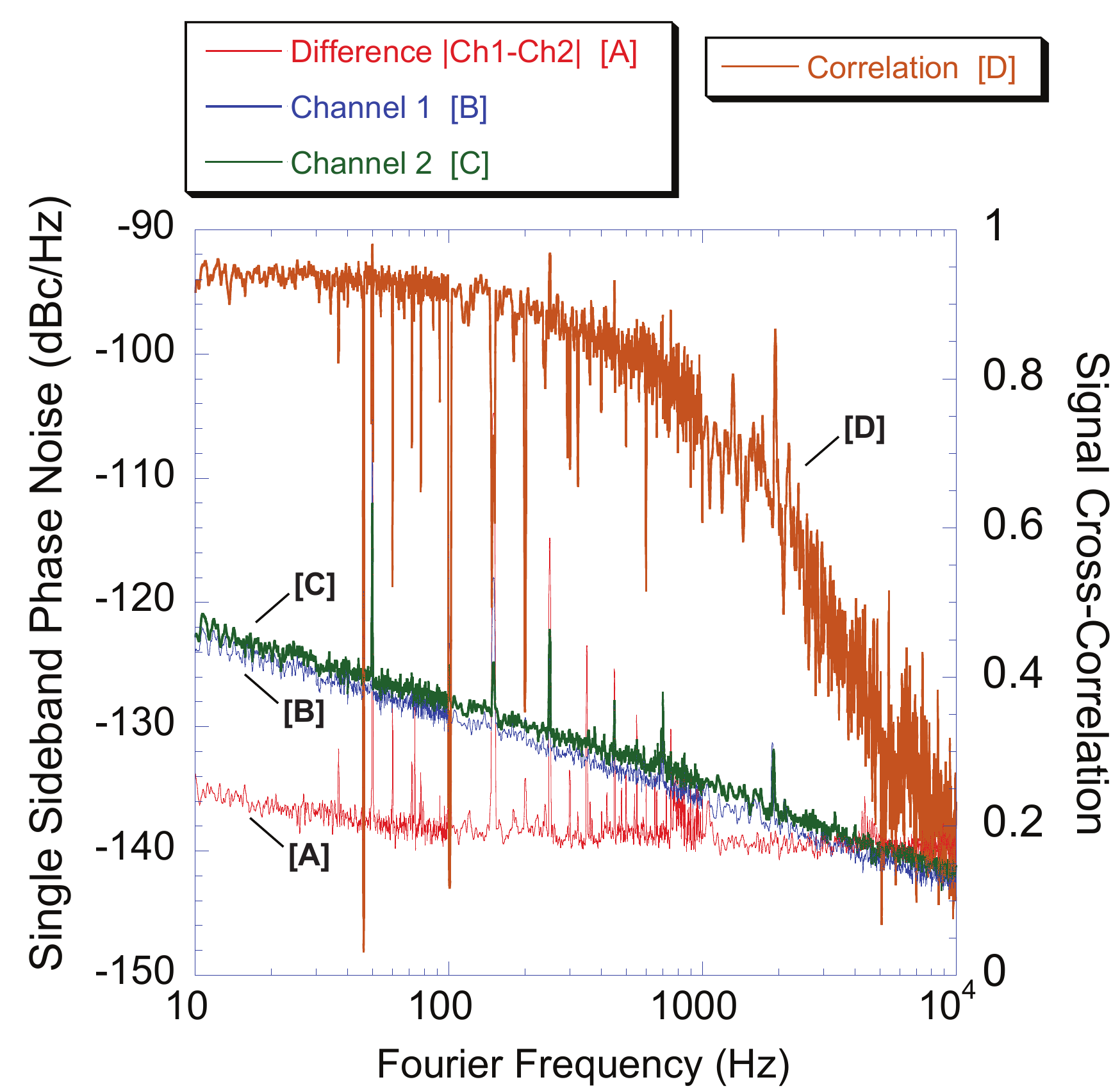}
\caption{\label{figure2}Phase Noise and signal cross-correlation from noise measurement system. Note that as the correlation is lost, the phase noise of the difference frequency becomes identical to that of the individual signals.}
\end{figure}

The experimental setup to measure amplifier phase noise for two jointly amplified signals (``dual frequency readout'') is shown in Fig. \ref{fig1}. The measurement is configured for a \hbox{two-channel} spectral noise measurement \cite{WallsFCS1988} and the coherence between the two channels is determined by using a spectrum analyser to measure the cross correlation function, with the \hbox{two-channel} readout mixers set phase-sensitive. The system can be calibrated by placing a voltage controlled phase shifter (VCP) at the input of the amplifier (not shown in Fig. \ref{fig1}). The microwave signals are generated from a fixed frequency low noise sapphire oscillator \cite{IvanovMTT1998,IvanovUFFC1998, IvanovTAP2000,IvanovMTT2006,IvanovUFFC2009} and a HP 8673G Signal Generator. The signals are offset slightly in frequency, with a mean frequency close to 9 GHz.
To create the dual frequency, the two microwave signals are power combined with a 3 dB hybrid and sent to a common amplifier. Once amplified, the dual frequency signal is split by a 3 dB hybrid and sent to two separate microwave mixer RF ports.  Each channel of the measurement system in Fig. \ref{fig1} is tuned phase sensitive by adjusting phase delay of the mixer LO signal. In such a case, voltage fluctuations at the output of each mixer are expected to vary synchronously with fluctuations of phase introduced by the microwave amplifier at the respective carrier frequency. The demodulated signals thus give a downconverted response of the amplifier with respect to the two initial frequencies. The demodulated signals are filtered and amplified using a Stanford Research Systems low noise amplifier with a bandwidth of 30 kHz, in order to deal with the low frequency noise only, and a gain of 20 dB. This increases the sensitivity of the phase noise measurement and filters the second frequency from the output to avoid saturation effects. In this measurement, the dual frequency was created with a 1.2 MHz frequency difference by setting the signal generator to 9.00078 GHz. The phase sensitivity was measured to be approximately 200 mV/rad for both channels. A two-channel FFT Spectrum Analyzer measured the outputs of the mixer so that the noise at both frequencies and the correlation between the noise could be determined. The results are given in Fig. \ref{figure2}, where the cross-correlation between the two channels is shown along with the single sideband phase noise measured for each channel as shown in Fig. \ref{fig1} and their difference. From the results in Fig. \ref{figure2} it is seen that at Fourier frequencies below 1 kHz, voltage fluctuations at both outputs of the measurement system are strongly correlated and therefore cancel each other upon subtraction. This leaves behind voltage noise with ``white'' spectrum resulting from thermal phase fluctuations of the microwave amplifier. At higher Fourier frequencies, performance is limited by the uncorrelated white phase noise floor.

\section{Effect of Amplifier Readout Noise on Frequency Stability}
In addition to the phase noise measurement presented in Section \ref{section2}, the performance of our `dual frequency' readout system was assessed in terms of fractional frequency instability. To test the two-oscillator beat frequency measurement technique at ultra-low levels of power and with world-leading stability, two nominally identical cryogenic sapphire oscillators (CSOs) close to 11.2 GHz were employed with heavily attenuated output signals. The two CSOs used here exhibit a relative fractional frequency instability of $1.6 \times 10^{-15}$ at 1 second and reach a minimum of $8 \times 10^{-16}$ at about 20s of integration before rising again to $1.4 \times 10^{-15}$ at 100s \cite{HartnettIFCS09}. The oscillators make use of a pair of HEMEX sapphire resonators from Crystal Systems (USA), operating in the quasi-transverse magnetic Whispering Gallery mode, WGH$_{16,0,0}$ with a frequency-temperature annulment point \cite{JonesBlair1988el,Hartnet1999uffc, Tobar1997jpDap} a few kelvin above liquid helium temperature. Two nominally identical loop oscillators were constructed. The parameters of each CSO are listed in Table \ref{Table1}. The beat frequency between the CSOs of 336 kHz falls within the most sensitive regime of the Agilent 53132A frequency counter, which was used in the readout system.
\begin{table}[h]
\renewcommand{\arraystretch}{1.3}
\caption{Parameters for the Cryogenic Sapphire Oscillator ensemble. $Q_L$ is the loaded Q-factor of the resonator, and $\beta_1$ \& $\beta_2$ are the couplings on ports 1 and 2 respectively.}
\label{Table1}
\centering
\begin{tabular}{|c|c|c|}
\hline
 & \textbf{CSO1} & \textbf{CSO2}\\
\hline
\hline
Frequency	& 11.200,386,540 GHz 	&	11.200,053,848 GHz\\
Temperature 	& 7.2319 K 				&	6.6179 K\\
$Q_L$		& $7 \times 10^8$		&	$4.4 \times 10^8$\\
$\beta_1$	& 1.06					&	0.61\\
$\beta_2$	& 0.1					&	0.035\\
\hline
\end{tabular}
\end{table}
\subsection{Measurement Technique}
\begin{figure*}[!t]
\includegraphics[width=7.2in]{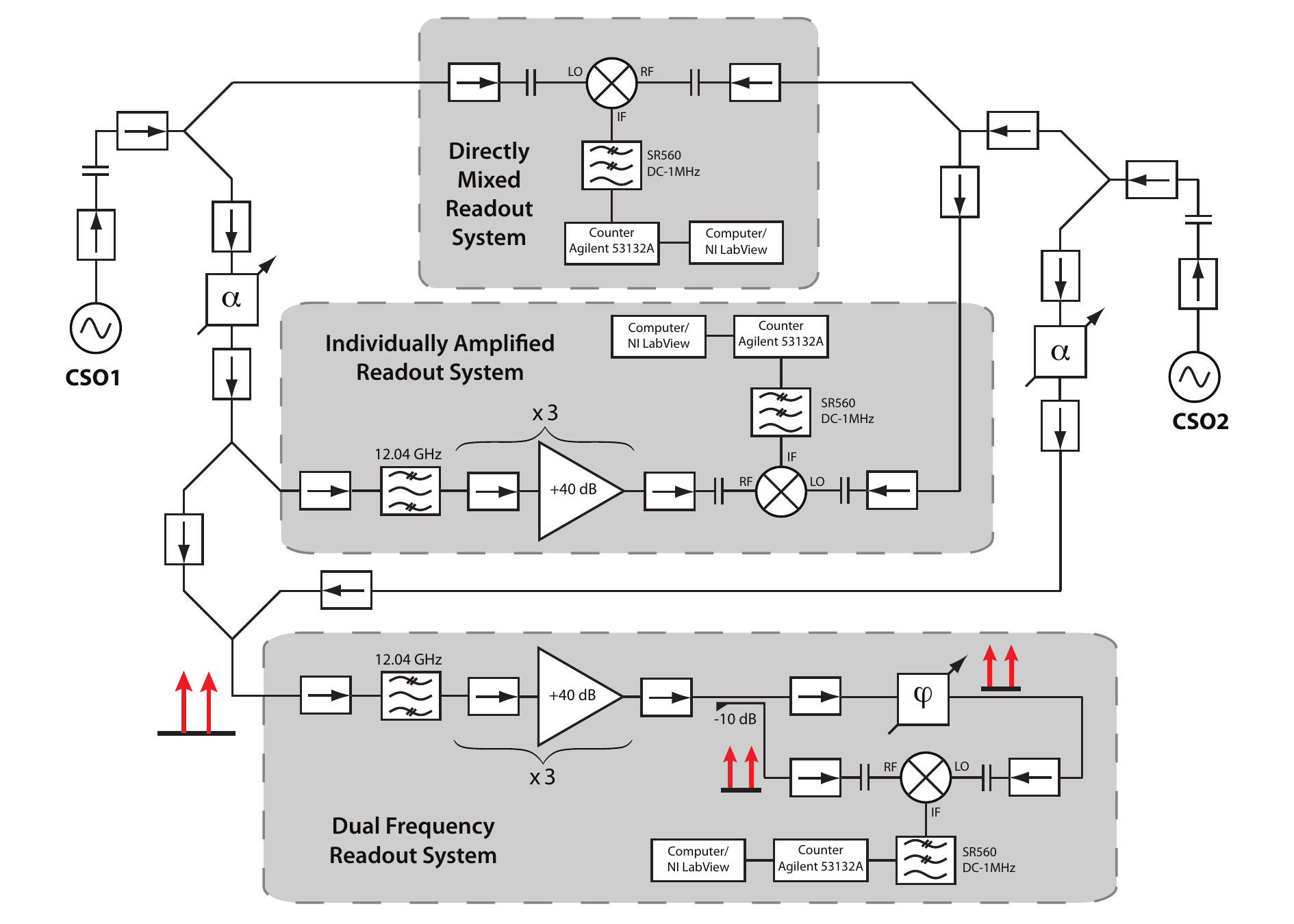}
\caption{\label{figure3}Schematic of the three simultaneous measurements of the three read out systems. The isolators used had a nominal insertion loss of typically $<0.2$dBm which was adjusted for in the variable attenuation.}
\end{figure*}

\begin{figure}[t]
\includegraphics[width=3.3in]{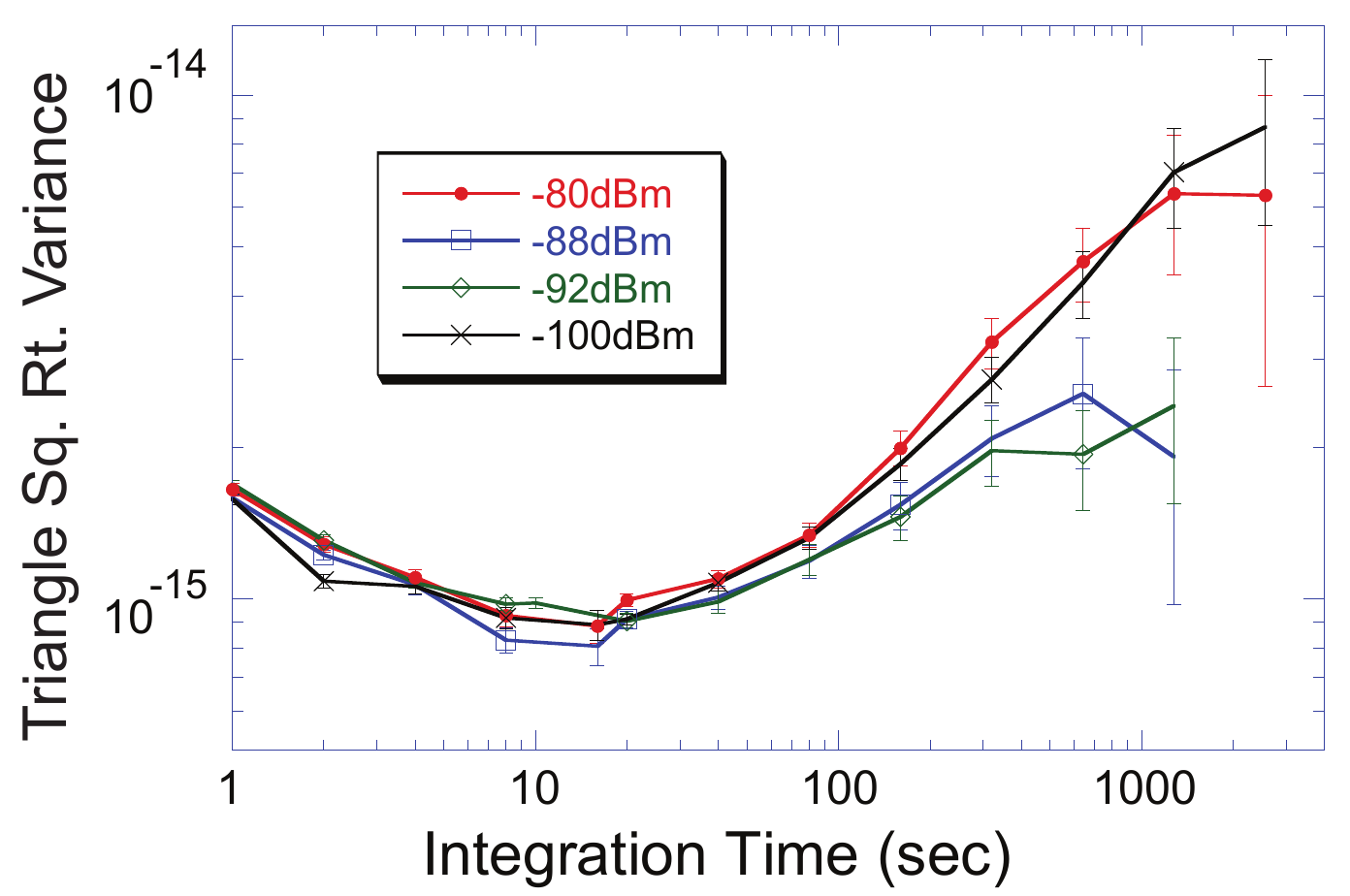}
\caption{\label{figure4}Triangle square root variance of the beat frequency for the directly mixed signals. This constitutes a `noise floor' for the other measurements.}
\end{figure}

Three readout systems were devised and tested simultaneously (Fig. \ref{figure3}) so that each system experienced identical conditions such as room temperature fluctuations, vibrational disturbances, oscillator drifts, and the inherent resonator and oscillator noise. The `control' or `directly mixed' system contained no amplifiers. The output signal from the loop oscillator of each CSO (+2 dBm for CSO1 and +17 dBm for CSO2) was first split using a power divider, and one output from each divider was passed directly into a \textit{Marki Microwave} M1R0818LS mixer without amplification or filtering of any kind. In all three systems, the input power for the mixer was adjusted for the optimal operating regime according to the mixer datasheet. The relative frequency instability of the resultant beat frequency gives the noise floor for the other measurement systems. Flicker phase noise added by the amplification chains in the other readout systems will manifest as a degradation of the fractional frequency stability. Great care was taken to provide adequate isolation at pertinent locations in the microwave circuits such as before and after amplifiers, on the input ports of mixers, and particularly after the entry point of the CSO signals. The inputs for the directly mixed readout system remain unchanged, so we expect the fractional frequency stability to remain constant in each test. However despite using 60 dB of isolation, some level of interference or `crosstalk' between signals is unavoidable, which is apparent at short integration times as seen in Fig. \ref{figure4}.\\

The `individually amplified' system is the standard readout system for two low-power signals. Each signal is passed through its own separate amplification chain before mixing. In this implementation, the second branch of the power divided signal from CSO1 is attenuated to simulate a low power ultra-stable signal. This signal is then amplified by a bank of three \textit{JCA} 812-4134 microwave amplifiers connected in series and passed to the RF port of a \textit{Marki Microwave} M1R0818LS mixer. The LO signal is taken from an unattenuated branch of split power from CSO2. The measured frequency instability of the beat signal then allows us to determine the effects of noise added by the amplification chain.\\

Finally, the self-mixed or `dual frequency readout' system combines the attenuated signals from both CSOs using a power divider in its reverse mode of operation. The two signals are then simultaneously passed through an amplification chain before being split with a directional coupler, phase adjusted, and self mixed as in Fig. \ref{figure3}. The RF and LO voltages may be written as:

\begin{equation}
\label{eqn1}
u_{\text{RF}}=U_{\text{RF}}\left(\text{Cos}\left[\omega _1t+\phi _1\right]+\text{Cos}\left[\omega _2t+\phi _2\right]\right)
\end{equation}%

\begin{equation}
\label{eqn2}
u_{\text{LO}}=U_{\text{LO}}\left(\text{Cos}\left[\omega _1t+\phi _1+\varphi\right]+\text{Cos}\left[\omega _2t+\phi _2+\varphi\right]\right)
\end{equation}

The optimal self-adjustment is calculated by assuming the mixer is an ideal multiplier, with the beat signal $u_B$ at the IF port proportional to $U_{\text{RF}}$ and $\text{Cos}[\Phi]$:

\begin{equation}
\label{eqn3}
u_{\text{B}} \approx U_{\text{RF}}{\text{ }}\text{Cos}[\varphi ]\text{ Cos}\left[\left(\omega _2-\omega _1\right)t+\left(\phi _2-\phi _1\right)\right]
\end{equation}

Here $\varphi$ is the phase in the self-mixed phase bridge in the Dual Frequency Readout of Fig. \ref{figure3}. Thus to maximize the signal into the frequency counter (lowering the noise floor) the phase is set to an integer number of $\pi$. For this setup, correlated noise introduced in the amplification chain is subtracted in the mixing stage.
In all cases the beat frequency is filtered and amplified with a Stanford Research Systems SR560 low noise preamplifier. The attenuation in each system was adjusted so that the power of the signal at the input of the amplifiers and mixers was identical. Variable mechanical attenuators were found to exhibit strong vibration sensitivity and were discarded in favor of fixed coaxial attenuators. The entire microwave circuit was bolted securely to an optical bench and the experiment was conducted in a room with temperature stability of $\pm 1^\circ$C.
Data was taken at four power levels: $-$80 dBm, $-$88 dBm, $-$93 dBm and $-$100 dBm. Attenuation was varied by using combinations of attenuators. At each power level, the beat frequency output data was collected for several hours with an Agilent 53132A frequency counter and logged on a computer running the NI Labview software. The beat frequency was measured for counter gate times of 1 second, 4 seconds, and 10 seconds, giving 16 sets of data in total.%
\subsection{Beat Note Phase Fluctuations}

In this section we use a simple model to analyze the difference between the dual frequency (DF) and individually amplified (IA) readout system in Fig. \ref{figure3}. The double sideband spectral density of amplifier phase noise is given by

\begin{equation}
\label{eqn4}
S_{\phi }^{\text{amp}}=\frac{\alpha }{f}+S_{\phi }^{\text{thermal}}\text{.}
\end{equation}

Here, the first term characterizes $1/f$ phase noise while the second term represents thermal noise fluctuations with spectral density:
\begin{equation}
\label{eqn5}
S_{\phi }^{\text{thermal}}=\frac{k_BT_0N}{P_{\text{inp}}}
\end{equation}
where $k_B$ is the Boltzmann constant, $T_0$ is the ambient temperature, $N$ is the amplifier noise figure and $P_{\text{inp}}$ is the power of the input signal. The parameter $\alpha$ generally takes a value between $10^{-10}$ and $10^{-12}$ and is independent of input power, as long as the amplifier is not driven into saturation. Assuming all amplifiers are identical, and that the dual frequency system has 100\% correlated flicker phase noise at both frequencies, the spectral density of noise introduced by the amplifiers in the dual frequency system is:

\begin{equation}
\label{eqn6}
S_{\phi}^{\text{DF}}=\frac{2k_BT_0N}{P_{\text{inp}}}
\end{equation}
while the noise in the individually amplified system is the same as that given in (\ref{eqn4}). However, for a fair comparison we need to assume both the LO and RF have similar levels so that:

\begin{equation}
\label{eqn7}
S_{\phi}^{\text{IA}}=\frac{2\alpha }{f}+\frac{2k_BT_0N}{P_{\text{inp}}}\text{.}
\end{equation}

In terms of the time domain, we can transform the white phase and flicker noise terms in (\ref{eqn7}) with respect to Triangle square root variance \cite{DawkinsTriangle2007} using (\ref{eqn8}) and (\ref{eqn9}) respectively.

\begin{equation}
\label{eqn8}
\sigma _{T}(\tau )=\frac{2}{\pi  f_0}\sqrt{\frac{k_BT_0N}{P_{\text{inp}}}}\text{ }\tau ^{-3/2}
\end{equation}%

\begin{equation}
\label{eqn9}
\sigma _{T}(\tau )=\frac{2}{\pi  f_0}\sqrt{3\alpha  \ln [27/16]}\text{ }\tau ^{-1}
\end{equation}

\begin{figure}[t]
\includegraphics[width=3.3in]{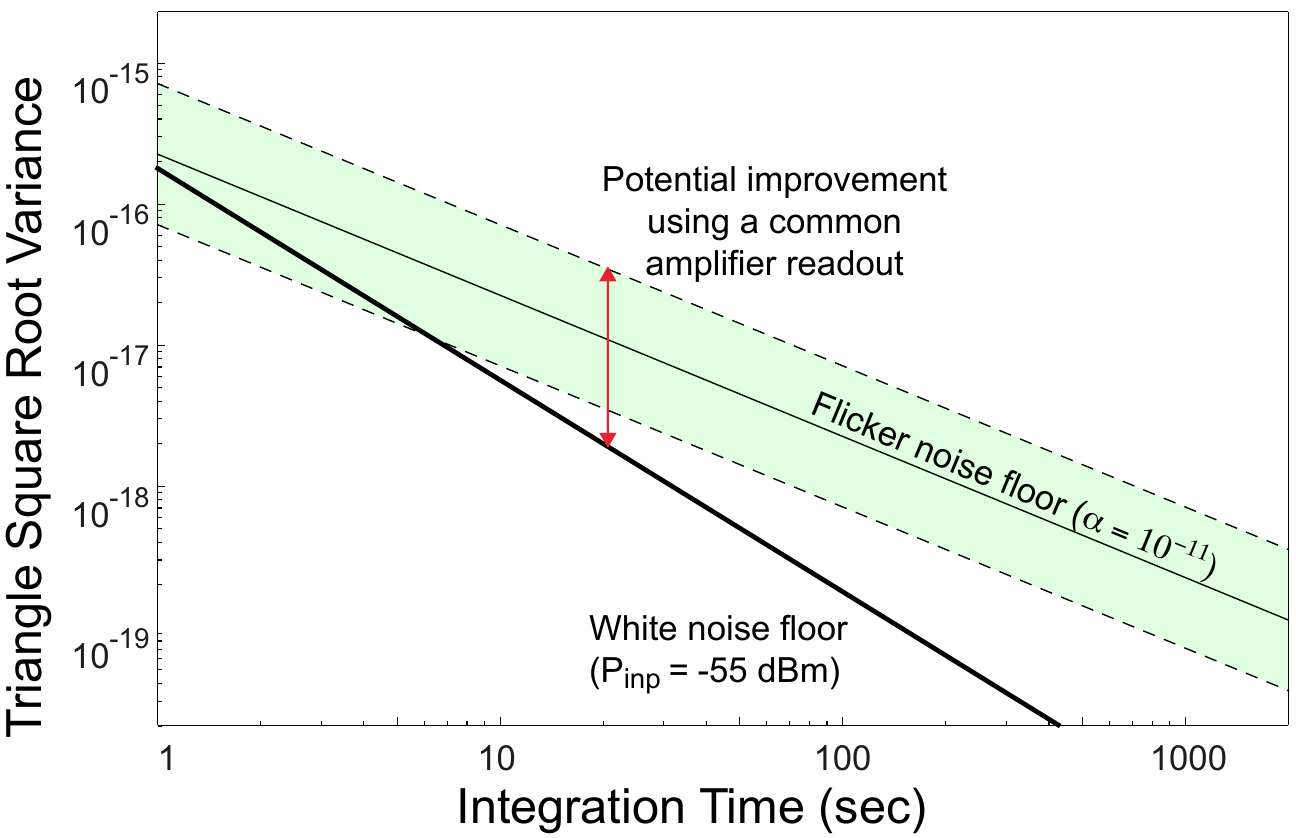}
\caption{\label{potentialimprovement}An illustrative example of the noise floors given by (\ref{eqn8}) and (\ref{eqn9}), calculated for a signal with an arbitrary $-55$ dBm power. The shaded area around the flicker noise floor shows the effect of varying the parameter $\alpha$ between $10^{-10}$ (upper noise floor limit) and $10^{-12}$ (lower noise floor limit). The white noise floor is dependent on $P_{\text{inp}}$, and becomes dominant over the flicker phase noise floor at very low power. Shown is the potential improvement in resolution when a common amplifier, flicker-noise-free readout system is used.}
\end{figure}

Here $f_0$ is the oscillator frequency. Considering the measured amplifier flicker noise in Fig. \ref{figure2} is $10^{-11}/f$  $\text{rad}^2/$Hz, we use this as a typical value to calculate the flicker noise floor with (\ref{eqn9}). Substituting $\alpha$ = $10^{-11}$ into (\ref{eqn9}) and given the frequency of the CSO is 11.2 GHz, the frequency instability due to the flicker phase noise in the readout system is $2.3 \times 10^{-16}/\tau$. This is below the noise floor provided by the oscillators presented here but close to the most stable oscillators built to date \cite{Hartenett2006apl,Locke2006rsi, HartnettNand2010,HartnettNandMTT2010}. Figure \ref{potentialimprovement} shows the potential improvement in measurement resolution by using a common amplifier readout system to cancel flicker phase noise. We have previously measured the fractional frequency resolution of a digital counter to be significantly less than the limits given in (\ref{eqn8}) and (\ref{eqn9}).\\

In principle, cryogenic sapphire oscillators and masers could be operated with performance at the thermal noise limit. This limit has already been measured at very low powers with frequency instabilities of order $10^{-14}$ \cite{Benmessai2008prl}. At medium power levels this limit can be well below $10^{-16}/\tau$, and at this level of performance, the flicker phase noise of the readout amplifier will need to be dealt with. The dual frequency readout presented in this work could be used to measure frequency instabilities below this level.\\

In the next section we show that the thermal noise of the readout amplifier limits the measurements of frequency instability at integration times below 10 seconds for power levels below $-$80 dBm. In addition, we show that the dual frequency readout is equivalent to the standard individually amplified  system in terms of added noise when at the thermal noise limit, but uses only half the number of amplifiers.

\subsection{Experimental Results}

The circuit as shown in Fig. \ref{figure3} was built to monitor and compare simultaneously the performance of the cryogenic oscillators and the three different readout systems. High-resolution Agilent frequency counters were used to monitor the beat frequency, which has been shown to naturally measure the Triangle variance \cite{DawkinsTriangle2007}. Thus, in this work we report the frequency deviation in terms of the Triangle variance.\\
\begin{figure}[b]
\includegraphics[width=3.3in]{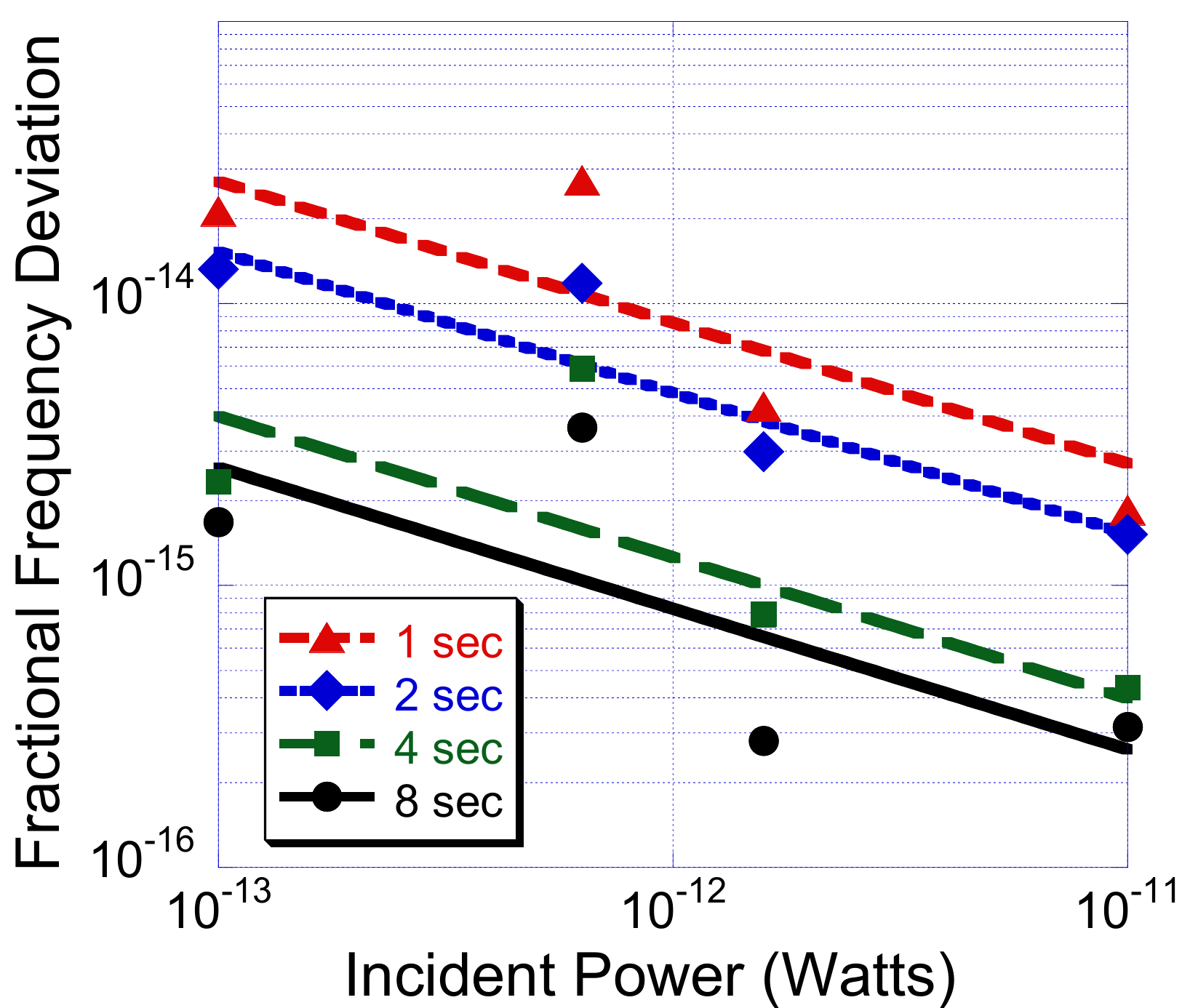}
\caption{\label{figure5}Frequency deviation (Triangle Square Root Variance) of the beat frequency for the individually amplified readout system.}
\end{figure}%

It should be noted that in the individually amplified system, only the signal from one CSO is attenuated then re-amplified, hence amplifier-induced noise is only added to one signal. Ordinarily, both signals would be amplified by separate amplification chains, however a lack of available amplifiers forced the measurement to be made this way. Compare this to the dual frequency system in which both signals are present in the amplification chain, hence noise was added to both signals. Thus, to determine an accurate measure of the noise introduced by each readout system, we subtract quadratically the noise floor (given by the directly mixed measurement shown in Fig. \ref{figure4}) from both the individually amplified and dual frequency measurements to give the residual noise of the readout. The noise in the individually amplified system is then multiplied by $\sqrt{2}$ to give a direct comparison to the dual frequency system. Data from the readout systems were composed of data points from the 1 second gate time ($\tau=$ 1 and 2 sec data points), 4 second gate time ($\tau=$ 4 and 8 sec data points), and 10 second gate time ($\tau\ge$ 10 sec data points).\\

\begin{figure}[t]
\includegraphics[width=3.3in]{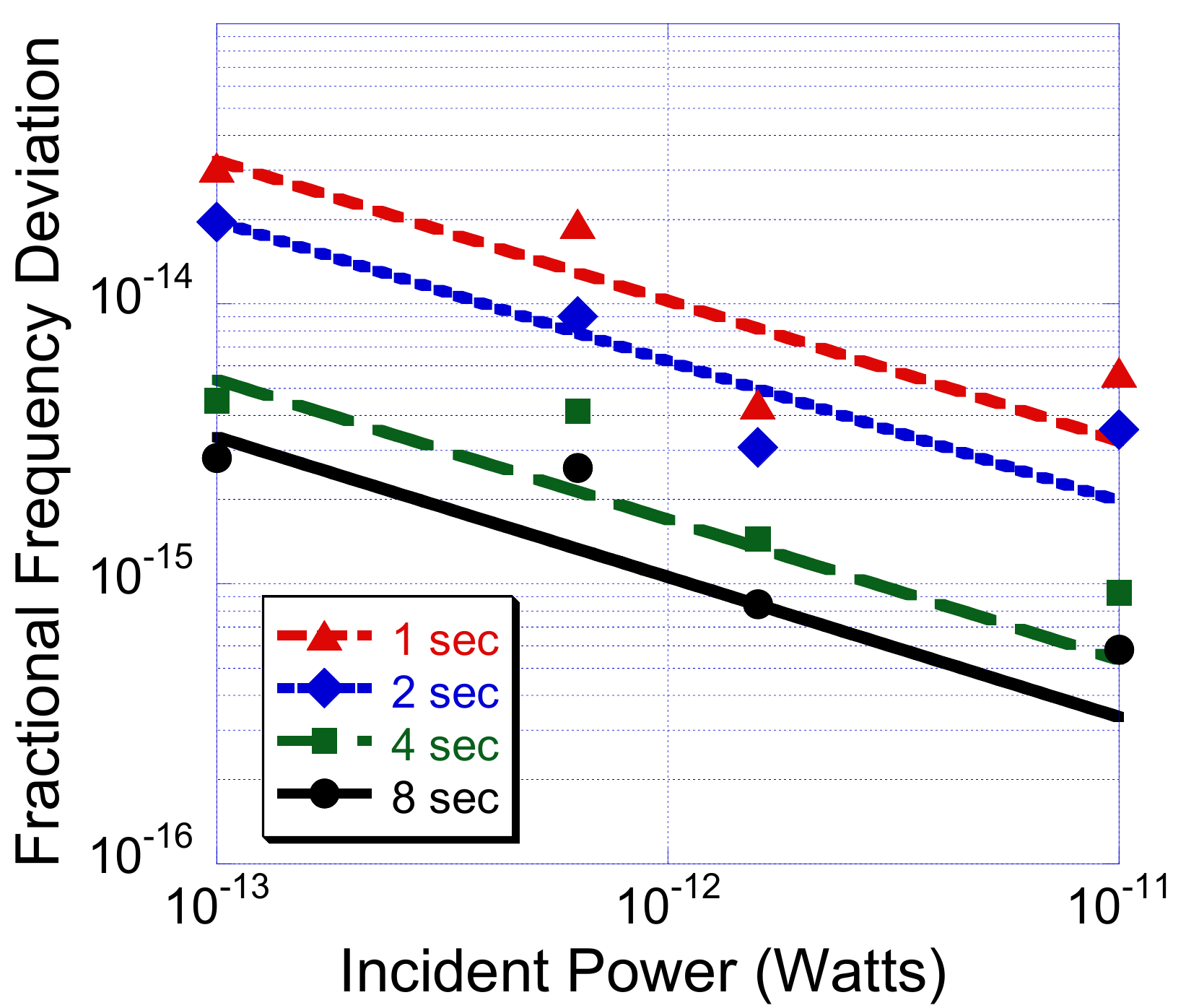}
\caption{\label{figure6}Frequency deviation (Triangle Square Root Variance) of the beat frequency for the dual frequency readout scheme.}
\end{figure}

\begin{figure}[t]
\includegraphics[width=3.3in]{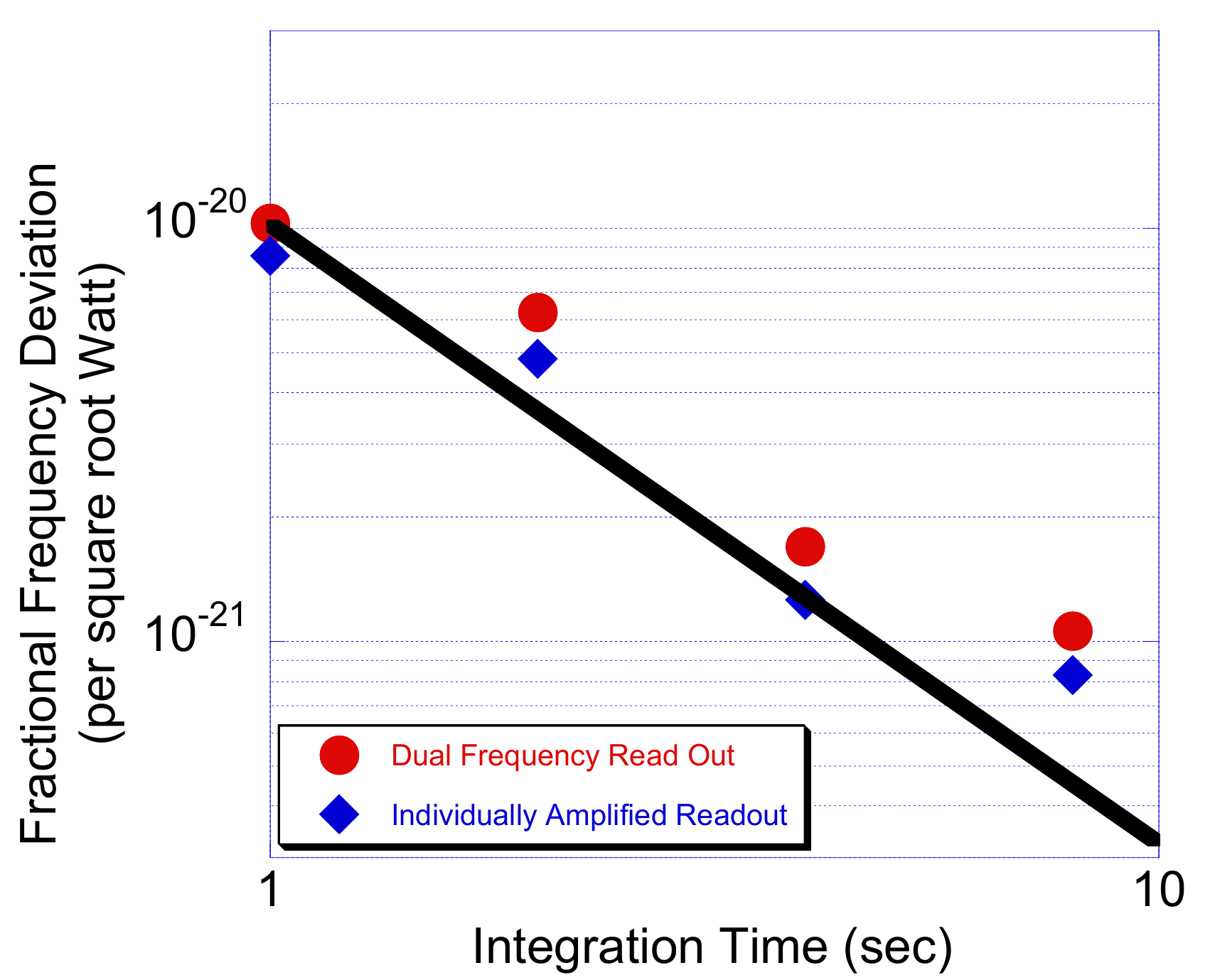}
\caption{\label{figure7}The fractional frequency deviation (Triangle Square Root Variance) per square root Watt as a function of integration (measurement) time. For short measurement times the noise shows a $\tau^{-3/2}$ dependence, which is the characteristic of added white phase noise by the readout amplifiers.}
\end{figure}

The residual noise of the individually amplified and dual frequency systems as a function of input power is shown for 1, 2, 4 and 8 second measurement times in Fig. \ref{figure5} and \ref{figure6} respectively. The curve fits reveal that the noise scales with power as thermal noise in (\ref{eqn8}) due to the $1/\sqrt{P_\text{inp}}$ dependence. Following this, we plot the fit coefficients from Fig. \ref{figure5} and \ref{figure6} as a function of measurement time, with units of fractional frequency deviation per square root Watt (Fig. \ref{figure7}). The dependence is close to $\tau^{-3/2}$ and the Individually Amplified and Dual Frequency results are within 20\% of each other. The dependence is approximately $10^{-20}\times\tau^{-3/2}$ and from (\ref{eqn8}) one can calculate the average noise figure of the amplifiers in the readout, which is equivalent to a value of 4dB. In these two-oscillator frequency instability measurements, at measurement times between 1 to 10 seconds, the noise was limited primarily by uncorrelated white phase noise for input power below $-$80 dBm and was not cancelled due to its non-correlated nature. 


\section{Conclusion}

We have shown that $1/f$ phase noise added by an amplifier operating in a `dual frequency' mode of excitation is correlated for both signals and is cancelled when a beatnote at the difference frequency is generated in a nonlinear mixing stage. In such a case, resolution with which the frequency of the beatnote can be measured is primarily limited by amplifier thermal fluctuations. The dual frequency readout system based on a common amplifier enables flicker noise free measurements of frequency fluctuations in high spectral purity weak microwave signals produced, for example, by solid-state cryogenic masers. In the white noise limited regime, our readout system performs as well as the conventional system using only half the number of microwave amplifiers.

\section*{Acknowledgment}
This work was funded by the Australian Research Council.

\bibliographystyle{IEEEtran}
\bibliography{IEEEabrv,biblio}

\begin{IEEEbiography}[{\includegraphics[width=1in,height=1.25in,clip,keepaspectratio]{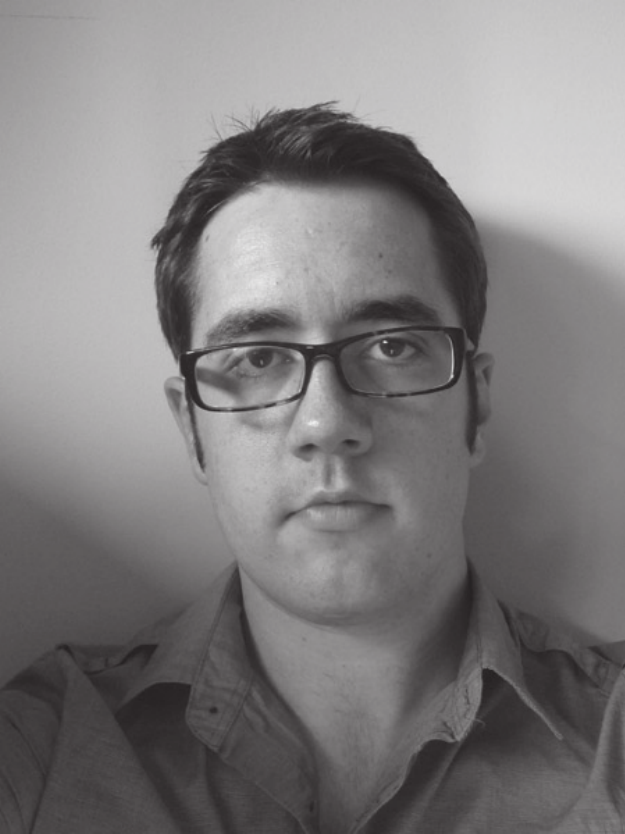}}]{Daniel L. Creedon}
was born in Perth, Western Australia in 1986. He received his B.Sc. (Hons.) degree in Physics at the University of Western Australia in 2007 and is now a Ph.D. student with the Frequency Standards and Metrology research group within the School of Physics. He is currently working on a cryogenic solid-state maser based on Electron Spin Resonance of paramagnetic ions in a sapphire Whispering Gallery mode resonator. He is a student member of the IEEE.
\end{IEEEbiography}
\begin{IEEEbiography}[{\includegraphics[width=1in,height=1.25in,clip,keepaspectratio]{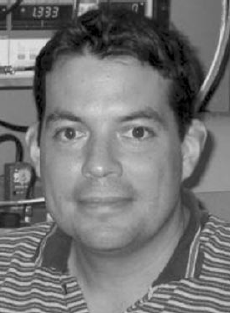}}]{Michael E. Tobar}
received the Ph.D. degree in physics from the University of Western Australia, Perth, W.A., Australia, in 1993. He is currently an ARC Laureate Fellow with the School of Physics, University of Western Australia. His research interests encompass the broad discipline of frequency metrology, precision measurements, and precision tests of the fundamental of physics. Prof. Tobar was the recipient of the 2009 Barry Inglis medal presented by the National Measurement Institute for precision measurement, the 2006 Boas medal presented by the Australian Institute of Physics, 1999 Best Paper Award presented by the Institute of Physics Measurement Science and Technology, the 1999 European Frequency and Time Forum Young Scientist Award, the 1997 Australian Telecommunications and Electronics Research Board (ATERB) Medal, the 1996 Union of Radio Science International (URSI) Young Scientist Award, and the 1994 Japan Microwave Prize. Professor Tobar is a Fellow of the IEEE.
\end{IEEEbiography}
\begin{IEEEbiography}[{\includegraphics[width=1in,height=1.25in,clip,keepaspectratio]{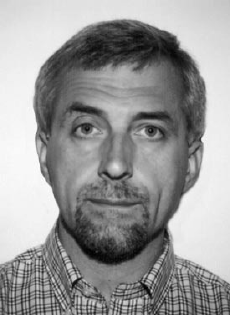}}]{Eugene N. Ivanov}received a PhD in Radiophysics from the Moscow Power Engineering Institute (MPEI) in 1987. In 1991, he joined the Physics Department at the University of Western Australia where he constructed a microwave readout system for the gravitational wave detector ``Niobe''. Since 1994 he has been working on applications of interferometric signal processing to generation of spectrally pure microwave signals and precision noise measurements. This research resulted in more than two orders of magnitude improvement in the phase noise of microwave oscillators relative to the previous state-of-the-art and enabled ``real'' time noise measurements with sensitivity exceeding the standard thermal noise limit. Since 1999, Eugene has worked as a Visiting Scientist at the National Institute of Standards and Technology (Boulder, Colorado). He identified and studied a number of noise mechanisms affecting fidelity of frequency transfer from the optical to the microwave domain. Dr. Ivanov is a recipient of the 1994 Japan Microwave Prize, 2002 W.G. Cady Award from the IEEE UFFC Society, and the 2010 J.F. Keithley Award from the American Physical Society.
\end{IEEEbiography}
\begin{IEEEbiography}[{\includegraphics[width=1in,height=1.25in,clip,keepaspectratio]{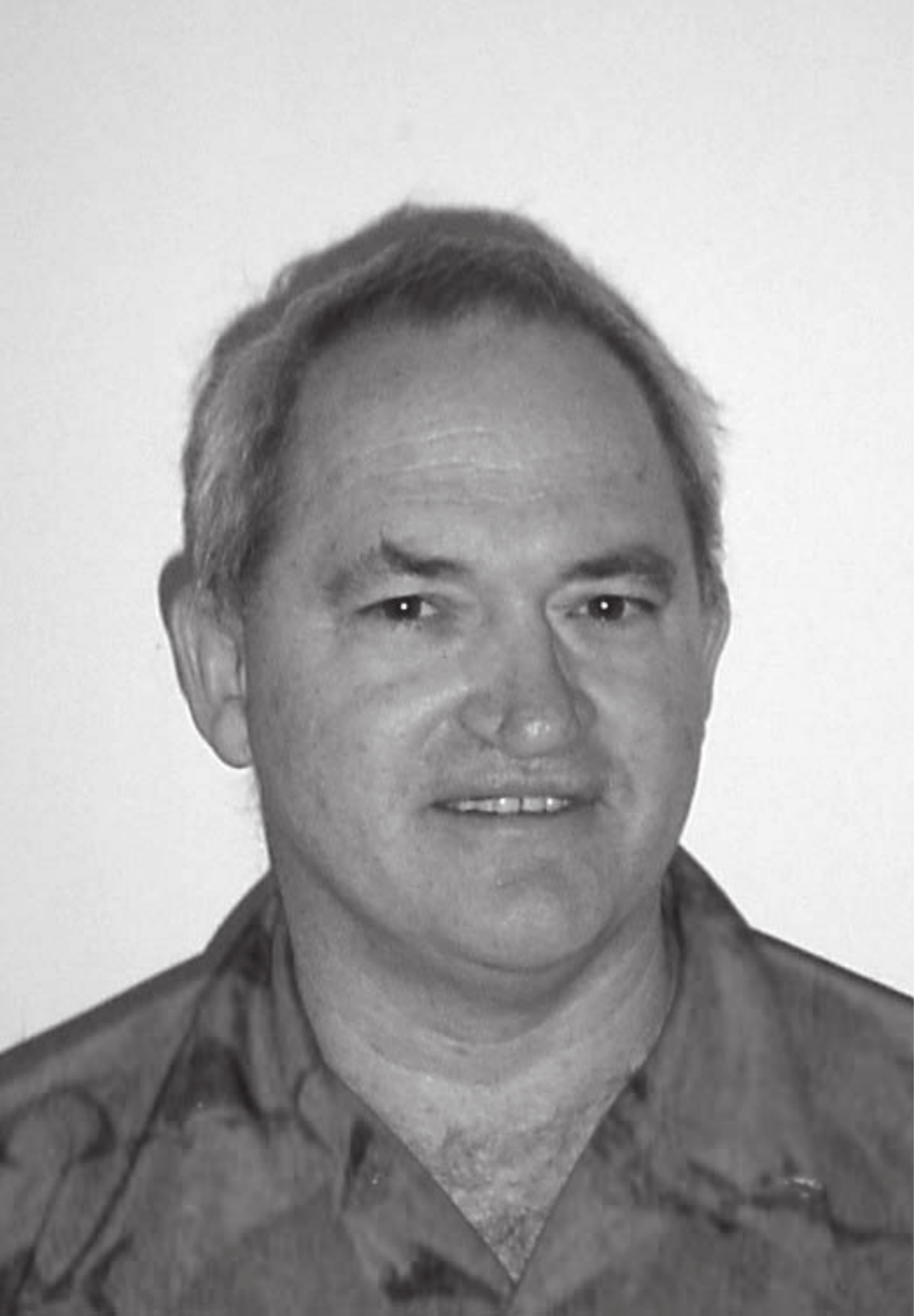}}]{John G. Hartnett}
received both his B.Sc. (1$^{\text{st}}$ class Hons.) and PhD (with Distinction) from the University of Western Australia (UWA). Prof. Hartnett was awarded the IEEE UFFC Society W.G. Cady award in 2010. He is co-recipient of the 1999 Best Paper Award presented by the Institute of Physics Measurement Science and Technology. He is currently a Research Professor with the Frequency Standards and Metrology research group (UWA). His research interests include planar metamaterials, the development of ultra-stable microwave oscillators based on sapphire resonators and tests of fundamental theories of physics such as Special and General Relativity using precision oscillators. 
\end{IEEEbiography}
\vfill

\end{document}